\documentclass[runningheads,a4paper]{llncs}

\usepackage{amssymb}
\usepackage{graphicx}
\usepackage{algorithm}
\usepackage{algpseudocode}
\usepackage{url}
\urldef{\mailsa}\path|{zhachen, hui.j.liu, yang6}@ucalgary.ca|
\newcommand{\keywords}[1]{\par\addvspace\baselineskip
\noindent\keywordname\enspace\ignorespaces#1}

\begin{document}

\mainmatter  

\title{Parallel Triangular Solvers on GPU}

\titlerunning{Parallel Triangular Solvers on GPU}

%
%
\author{Zhangxin Chen
\thanks{
The support of Department of Chemical and Petroleum Engineering, University of
Calgary and Reservoir Simulation Group is gratefully acknowledged. The research is
partly supported by NSERC/AIEE/Foundation CMG and AITF Chairs.}
 \and Hui Liu \and Bo Yang}
\authorrunning{Parallel Triangular Solvers on GPU}

\institute{University of Calgary\\
2500 University Dr NW, Calgary, AB, Canada, T2N 1N4\\
\mailsa}

%
%

\toctitle{Lecture Notes in Computer Science}
\tocauthor{Authors' Instructions}
\maketitle

\begin{abstract}
In this paper, we investigate GPU based parallel triangular solvers systematically. The parallel triangular solvers
are fundamental to incomplete LU factorization family preconditioners and algebraic multigrid solvers.
We develop a new matrix format suitable for GPU devices. Parallel lower triangular solvers and upper triangular solvers are
developed for this new data structure. With these solvers, ILU preconditioners and domain decomposition preconditioners are developed.
Numerical results show that we can speed triangular solvers around seven times faster.

\end{abstract}

\section{Introduction}
In many scientific applications, we need to solve lower triangular problems and upper triangular problems,
such as Incomplete LU (ILU) preconditioners, domain decomposition preconditioners and Gauss-Seidel
smoothers for algebraic multigrid solvers \cite{template,yousef2}.
The algorithms for these problems are serial in nature and difficult to parallelize on parallel platforms.
GPU is one of these parallel devices, which is powerful in float point calculation and is over 10 times faster
than latest CPU. Recently, GPU has been popular in various numerical scientific
applications. It is efficient for vector operations. Researchers have developed linear solvers for GPU devices
\cite{nv-spmv,nv-spmv2,hector1,liruipeng,cusp}. However, the development of efficient parallel triangular solvers for GPU is still
challenging \cite{cusp,lieb,nv-tri}.

Klie et al. (2011) investigated a triangular solver \cite{hector1}. They developed a level schedule method and a speedup of two was
obtained.  Naumov (2011) from the NVIDIA company also developed parallel triangular solvers \cite{nv-tri}.
He developed new parallel triangular solvers based on a graph analysis. The average speedup was also around two.

In this paper, we introduce our work on speeding triangular solvers. A new matrix format, HEC (Hybrid ELL and
CSR), is developed. A HEC matrix includes two matrices, an ELL matrix and a CSR
matrix. The ELL part is in column-major order and is designed the way to
increase the effective bandwidth of NVIDIA GPU. For the CSR matrix, each row
contains at least one non-zero element. This design of the CSR part reduces the
complexity of the solution of triangular systems. In addition, parallel algorithms for
solving the triangular systems are developed. The algorithms are motivated by the level schedule
method described in \cite{yousef2}. Our parallel triangular solvers can be sped up to seven times faster.
Based on these modified algorithms, ILU(k), ILUT and domain decomposition (Restricted
Additive Schwarz) preconditioners are developed. Numerical experiments are performed. These experiments show that we
can speed linear solvers around ten times faster.

The layout is as follows. In \S 2, a new matrix format and algorithms for lower triangular problems and upper triangular problems
are introduced. In \S 3, parallel triangular solvers are employed to develop ILU preconditioners and domain decomposition
preconditioners. In \S 4, numerical tests are performed. At the end, conclusions are presented.

\section{Parallel Triangular Solvers}

For the most commonly used preconditioner ILU, the following triangular
systems need to be solved:
\begin{equation}
\label{equ-lu} L U x = b \Leftrightarrow L y = b, U x = y,
\end{equation}
where $L$ and $U$ are a lower triangular matrix and an upper triangular matrix,
respectively, $b$ is the right-hand side vector, $x$ is the unknown to be solved for,
and $y$ is the intermediate unknown.
The lower triangular problem, $L y = b$, is solved first, and then, by solving
the upper triangular problem, $U x = y$, we can obtain the result $x$. In this paper,
we always assume that each row of $L$ and $U$ is sorted in ascending order according to
their column indices.

\subsection{Matrix format}

\begin{figure}
    \centering
    \includegraphics[width=0.80\linewidth]{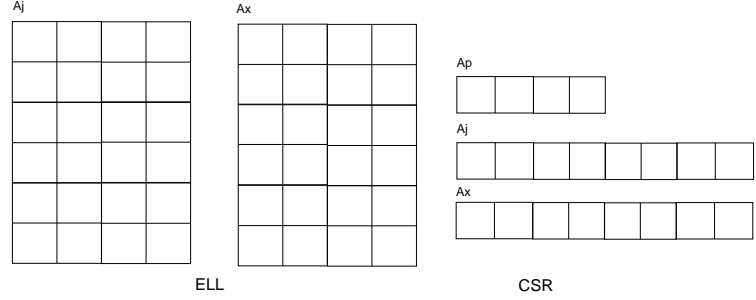}
    \caption{HEC matrix format.}
    \label{fig1}
\end{figure}

The matrix format we develop is denoted by HEC (Hybrid ELL and CSR) \cite{uc-spmv2}. Its basic
structure is demonstrated by Figure~\ref{fig1}. An HEC matrix contains two
submatrices: an ELL matrix, which was introduced in ELLPACK~\cite{ellpack}, and a CSR
(Compressed Sparse Row) matrix. The ELL matrix has two submatrices, a column-indices
matrix and a non-zeros matrix. The length of each row in these two matrices is
the same. The ELL matrix is in column-major order and is aligned when being
stored on GPU. Note that the data access pattern of global memory for NVIDIA
Tesla GPU is coalesced \cite{cuda-guide3} so the data access speed for the ELL
matrix is high. A disadvantage of the ELL format is that it may waste memory
if one row has too many non-zeros. In this paper, a CSR submatrix is added to overcome this problem. A CSR
matrix contains three arrays, the first one for the offset of each row, the second one
for column indices and the last one for non-zeros. For our HEC format matrix,
we store the regular part of a given triangular matrix $L$ in the ELL part and
the irregular part in the CSR part.
When we store the lower triangular matrix, each row of the CSR matrix has at least one
element, which is the diagonal element in the triangular matrix $L$.

\subsection{Parallel lower triangular solvers}
In this section, we introduce our parallel lower triangular solver to solve
\[
L x = b.
\]
The solver we develop is based on the level schedule method \cite{yousef2,liruipeng}.
The idea is to group unknowns $x(i)$ into different levels so that
all unknowns within the same level can be computed simultaneously
\cite{yousef2,liruipeng}. For the lower triangular problem, the level of
$x(i)$ is defined as
\begin{equation}
\label{equ-level}
 l(i) = 1 + \max_j {l(j)} \quad {\ for \ all\ } j {\ such\
that\ } \ L_{ij} \neq 0, i = 1, 2, \ldots, n,
\end{equation}
where $L_{ij}$ is the $(i, j)$th entry of $L$, $l(i)$ is zero initially and $n$ is the number of rows.

We define $S_i = \{x(j) : l(j) = i\}$, which is the union of all unknowns
that have the same level. Here we assume that each set $S_i$ is sorted  in ascending
order according to the indices of the unknowns belonging to $S_i$. Define $N_i$ the
number of unknowns in set $S_i$ and $nlev$ the number of levels. Now, a map
$m(i)$ can be defined as follows:
\begin{equation}
\label{equ-map}
m(i) = \sum_{j = 1}^{k - 1} N_j + p_k(x(i)), \quad x(i) \in S_k,
\end{equation}
where $p_k(x(i))$ is the position $x(i)$ in the set $S_k$ when $x(i)$ belongs
to $S_k$. With the help of the map, $m(i)$, we can reorder the triangular matrix
$L$ to $L'$, where $L_{ij}$ in $L$ is transformed to $L'_{m(i)m(j)}$ in $L'$.
$L'$ is still a lower triangular matrix. From this map, we find that if $x(i)$
is next to $x(j)$ in some set $S_k$, the $i$th and $j$th rows of $L$ are next
to each other in $L'$ after reordering. It means that $L$ is reordered level by
level, which implies that memory access in matrix $L'$ is less irregular than that
in matrix $L$. Therefore, $L'$ has higher performance compared to $L$ when we
solve a lower triangular problem.

The whole algorithm is described in two steps, the preprocessing step and the solution
step, respectively. The preprocessing step is described in Algorithm \ref{alg1}.
In this step, the level of each unknown is calculated first. According to these
levels, a map between $L$ and $L'$ can be set up according to equation
(\ref{equ-map}). Then the matrix $L$ is reordered. We should mention that $L$ can
be stored in any kind of matrix format. A general format is CSR. At the end,
$L'$ is converted to the HEC format and as we discussed above each row of the CSR
part has at least one element.

\begin{algorithm}
\caption{Preprocessing lower triangular problem}
\label{alg1}
\begin{algorithmic}[1]
\State {Calculating the level of each unknown $x(i)$ using equation (\ref{equ-level})}.

\State {Calculating the map $m(i)$ using equation (\ref{equ-map}).}

\State {Reordering matrix $L$ to $L'$ using map $m(i)$.}

\State {Converting $L'$ to the HEC format.}

\end{algorithmic}

\end{algorithm}

\begin{algorithm}
\caption{Parallel the lower triangular solver on GPU, $Lx = b$} \label{alg2}
\begin{algorithmic}[1]
\For {i = 1: n} \Comment{Use one GPU kernel to deal with this loop}
  \State $b'(m(i)) = b(i)$;
\EndFor

\For {i = 1 : nlev} \Comment {$L'x'=b'$}
  \State start = level(i);
  \State end = level(i + 1) - 1;
  \For {j = start: end} \Comment{Use one GPU kernel to deal with this loop}
  \State solve the $j$th row;
  \EndFor
\EndFor

\For {i = 1: n} \Comment{Use one GPU kernel to deal with this loop}
  \State $x(i) = x'(m(i))$;
\EndFor
\end{algorithmic}
\end{algorithm}

The second step is to solve the lower triangular problem. This step is
described in Algorithm \ref{alg2}, where $level(i)$ is the start row position
of level $i$. First, the right-hand side $b$ is permutated according to the
map $m(i)$ we computed. Then the triangular problem is solved level by level
and the solution in the same level is simultaneous. Each thread is responsible for one row.
At the end, the final solution is obtained by a permutation.

To solve the upper triangular problem $U x = b$, we introduce the following transferring map:

\[
t(i) = n - i,
\]
where $n$ is the number of rows in $U$. With this map, the upper triangular problem is transferred to
a lower triangular problem, and the lower triangular solver is called to solve the problem.

\section{Preconditioners}
The ILU factorization for a sparse matrix $A$ computes
a sparse lower triangular matrix $L$ and a sparse upper triangular
matrix $U$. If no fill-in is allowed, we obtain the so-called ILU(0)
preconditioner. If fill-in is allowed, we obtain the ILU(k) preconditioner, where $k$ is the fill-in level.
Another method is ILUT(p,tol), which drops entries based on
the numerical values $tol$ of the fill-in elements and the maximal number $p$ of non-zeros in each row \cite{yousef2,liruipeng}.

The performance of parallel triangular solvers for ILU(k) and ILUT is dominated by original problems. In this paper, we implement
block ILU(k) and block ILUT preconditioners. When the number of blocks is large enough, we will have sufficient parallel
performance. If the matrix is not well partitioned and the number of blocks is too large, the effect of ILU(k) and ILUT will be
weakened. When we partition a matrix, the graph library METIS \cite{metis} is employed.

As we discussed above, the effect of ILU preconditioners is weakened if the number of blocks is too large.
The domain decomposition preconditioner is implemented, which was developed by Cai et al. \cite{cai}.
The domain decomposition preconditioner we implement is the so-called Restricted Additive Schwarz (RAS) method.
Overlap is introduced, and, therefore, this preconditioner is not as sensitive as block ILU preconditioners.
It can lead to good parallel performance and good preconditioning.
In this paper, we treat the original matrix as an
undirected graph and this graph is partitioned by METIS \cite{metis}. The
subdomain can be extended according to the topology of the graph. Then each smaller
problem can be solved by ILU(k) or ILUT. In this paper, we use ILU(0).

Assume that the domain is decomposed into $s$ subdomains, and we have $s$ smaller problems, $A_1$, $A_2$, $\cdots$, and $A_s$.
We do not solve these problems one by one but we treat them as one bigger problem:
\[
diag(A_1, A_2, \cdots, A_s) x = b.
\]
Each $A_i$ is factorized by the ILU method, where we have
\begin{equation}
\label{ddm}
diag(L_1, L_2, \cdots, L_s) \times  diag(U_1, U_2, \cdots, U_s) x = L \times U x = b.
\end{equation}
Then equation (\ref{ddm}) is solved by our triangular solvers.

\section{Numerical Results}
In this section, numerical experiments are presented, which are performed on
our workstation with Intel Xeon X5570 CPUs and NVIDIA Tesla C2050/C2070 GPUs. The
operating system is CentOS 6.2 X86\_64 with CUDA Toolkit 4.1 and GCC 4.4. All
CPU codes are compiled with -O2 option.

The data type of a float point number is double. The linear solver is GMRES(20). BILU,
BILUT and RAS denote the block ILU(0), block ILUT and Restricted Additive Schwarz
preconditioners, respectively.

\begin{example} The matrix used in this example is from a three-dimensional
Poisson equation. The dimension is 1,000,000 and the number of
non-zeros is 6,940,000. The ILU preconditioners we use in this example are block ILU(0) and block
 ILUT(7, 0.1). The performance data is collected in Table \ref{tab-ex1}.
\end{example}

\begin{table}
\caption{Performance of the matrix from the Poisson equation} \label{tab-ex1}
\begin{tabular}{llllllll} \hline \noalign{\smallskip}
Pre & Blocks & CPU (s)   & GPU (s) & Speedup & Pre CPU (s)&Pre GPU (s) & Speedup \\
\noalign{\smallskip}
\hline
BILU     & 16  & 20.12  & 2.53  & 7.93  &0.0197 & 0.0057 & 3.46\\

BILU     & 128  &  20.2  & 2.39 & 8.44  & 0.0192 & 0.0055 & 3.46 \\

BILU     & 512  & 22.76  & 2.70  & 8.39 & 0.0188& 0.0052& 3.61 \\

BILUT     & 16 & 14.85  & 2.38  & 6.19  &  0.0241 & 0.010 & 2.27\\

BILUT     & 128  & 14.56  & 2.30  & 6.30 & 0.0239& 0.011 & 2.25 \\

BILUT     & 512 & 18.59  & 2.71  & 6.83 & 0.0234 & 0.010 & 2.24 \\

RAS & 256 & 18.70  & 2.16  & 8.60 & 0.0306 & 0.0051 & 5.94 \\

RAS & 2048 & 22.03  & 2.35  & 9.32 & 0.0390 & 0.0056 & 6.92\\

\hline

\end{tabular}

\end{table}

For the block ILU(0), we can speed it over 3 times faster. When the number of blocks increases, our algorithm has better
speedup. The whole solving part is sped around 8 times faster. BILUT is a special preconditioner, since it is obtained according to the
values of $L$ and $U$; sometimes, the sparse patterns of $L$ and $U$ are more
irregular than those in BILU. From Table \ref{tab-ex1}, the speedups are a little lower
compared to BILU. However, BILUT reflects the real data dependence, and its
performance is better in general. This is demonstrated by Table \ref{tab-ex1}.
The CPU version BILUT always takes less time than the CPU BILU. But for the GPU
versions, their performance is similar. The BILUT is sped around 2 times faster. The whole solving part is sped around 6 times faster.

The RAS preconditioner is always stable. It also has better speedup than BILU and BILUT. The triangular solver is sped around 6 times
 faster and the average speedup is around 9.

\begin{example}
Matrix atmosmodd is taken from the University of Florida sparse matrix
collection \cite{uf} and is derived from a computational fluid dynamics (CFD) problem.
The dimension of atmosmodd is 1,270,432 and it has 8,814,880 non-zeros. The ILU preconditioners we use in this example are block ILU(0) and block
 ILUT(7, 0.01). The performance data is collected in Table \ref{tab-ex2}.
\end{example}

\begin{table}
\caption{Performance data of the matrix atmosmodd} \label{tab-ex2}

\begin{tabular}{llllllll} \hline \noalign{\smallskip}
Pre & Blocks & CPU (s)   & GPU (s) & Speedup & Pre CPU (s)&Pre GPU (s) & Speedup \\
\noalign{\smallskip}
\hline

BILU     & 16 & 20.61  & 2.63  & 7.79  & 0.0248 & 0.0072 & 3.45\\

BILU     & 128 &  23.94  & 2.80 & 8.50 & 0.0244 & 0.0072 & 3.40\\

BILU     & 512 & 24.13  & 2.72  & 8.82 & 0.0241 & 0.0070 &  3.46 \\

BILUT     & 16 & 14.70  & 2.37  & 6.16  & 0.028669 & 0.0114 & 2.51 \\

BILUT     & 128 & 16.58  & 2.43  & 6.79 & 0.028380 & 0.0100 & 2.84 \\

BILUT     & 512 & 19.91  & 2.64  & 7.50 & 0.027945 & 0.0113 & 2.47 \\

RAS & 256 & 23.45  & 2.66  & 8.75 & 0.0428 & 0.0072 & 5.96\\

RAS  & 2048& 25.75  & 3.28  & 7.81 & 0.0546 & 0.0083 & 6.59 \\

\hline

\end{tabular}
\end{table}

The performance of BILU, BILUT and RAS is similar to that of the same preconditioners in Example 1. When BILU is
applied, the triangular solvers are sped over 3 times faster and the speedup of the whole solving phase is about 7.
Because of the irregular non-zero pattern of BILUT, the speedup of BILUT is around 2. The average speedup of the solving phase
is about 6. The speedup increases when the number of blocks grows. RAS is
as stable as in Example 1. The triangular solvers are sped around 6, and the average speedup of the solving phase is around 8.

\begin{example}
A matrix from SPE10 is used \cite{Chen-Huan-Ma}. SPE10 is a standard benchmark for the black oil
simulator. The problem is highly heterogenous and it is designed the way so
that it is hard to solve.
The grid size for SPE10 is 60x220x85. The number of unknowns is 2,188,851 and
the number of non-zeros is 29,915,573.
The ILU preconditioners we use in this example are block ILU(0) and block ILUT(14, 0.01).
The performance data is collected in Table \ref{tab-ex3}.
\end{example}

\begin{table}
\caption{Performance of SPE10} \label{tab-ex3}
\begin{tabular}{llllllll} \hline \noalign{\smallskip}
Pre & Blocks & CPU (s)   & GPU (s) & Speedup & Pre CPU (s)&Pre GPU (s) & Speedup \\
\noalign{\smallskip}
\hline

BILU     & 16 & 92.80  & 12.78  & 7.25  & 0.0421 & 0.0118 & 3.54 \\

BILU     & 128 &  86.22  & 12.05 & 7.14 & 0.0423 & 0.0119 & 3.56\\

BILU     & 512 & 92.82  & 12.87  & 7.20 & 0.0424 & 0.0119 & 3.56 \\

BILUT     & 16 & 32.00 & 7.17  & 4.46  & 0.0645 & 0.0747 & 0.86 \\

BILUT     & 128 & 42.51  & 7.82  & 5.42 & 0.0647 & 0.0747 & 0.86\\

BILUT     & 512 & 47.44  & 8.80  & 5.37 & 0.0645 & 0.0747 & 0.86 \\

RAS & 256 & 106.61  & 14.36  & 7.41 & 0.100 & 0.0198 & 5.07 \\

RAS & 1024 & 110.36  & 16.36  & 6.73 & 0.124 & 0.0242 & 5.11 \\
\hline

\end{tabular}
\end{table}

From Table \ref{tab-ex3}, we can speed the whole solving phase 6.2 times faster when ILU(0) is applied. The speedup increases
if we increase the number of blocks. The average speedup of BILU is about
3 and the average speedup for the whole solving stage is about 7. In this example, BILUT is the best, which always takes the least running time.
However, due to its irregular non-zero pattern, we fail to speed the triangular solvers. The
RAS preconditioner is always stable, and the average speedup of RAS is about 5, while the average speedup of the solving phase is around 6.5.

\section{Conclusions}
We have developed a new matrix format and its corresponding triangular solvers. Based on
them, the block ILU(k), block ILUT and Restricted Additive Schwarz preconditioners have been
implemented. The block ILU(0) is sped over three times faster, the block ILUT is sped around 2 times, and the RAS preconditioner
is sped up to 7 times faster. The latter preconditioner is very stable and it can be served as a general preconditioner for
parallel platform.

\bibliographystyle{latex8}

\end{document}